\documentstyle[preprint,aps,tighten,epsf,floats]{revtex}
\newcommand{\ben}{\begin{displaymath}}
\newcommand{\een}{\end{displaymath}}
\newcommand{\be}{\begin{equation}}
\newcommand{\ee}{\end{equation}}
\newcommand{\bea}{\begin{eqnarray}}
\newcommand{\eea}{\end{eqnarray}}

\newcommand{\fign}[1]{\label{#1}}

\begin{document}
\draft
\title{Matching Heavy Particle Approach to Relativistic Theory}
\author{ J. Gegelia${ }^a$\footnote{e-mail address:
gegelia@daria.ph.flinders.edu.au}   and  G.Japaridze${ }^b$ }
\address{${ }^a$ School of Physical Sciences, Flinders University of South 
Australia, \\ Bedford Park, S.A. 5042, Australia. \\ 
${ }^b$ Department of Physics, Centre for Theoretical studies of Physical 
systems, \\ Clark 
Atlanta University, Atlanta, GA 30314, U.S.} 
\date{\today}
\maketitle 

\begin{abstract}
On the simple model of interacting massless and heavy scalar fields it is
demonstrated 
that the technique of heavy baryon chiral perturbation theory reproduces the
results of relativistic theory. Explicit calculations are performed for
diagrams including two-loops. 
\end{abstract}

\section{introduction}
Heavy Baryon Chiral Perturbation Theory (HBCHPT), suggested in
\cite{jenkins} is an
important and effective method of calculation of different processes involving
electro-magnetic 
and strong interactions 
(For review and references see \cite{bernard},
\cite{fettes}). The authors of Ref. \cite{jenkins} used 
the ideas of the heavy 
quark effective field theory which allowed them to
avoid 
severe complications appearing in problem of relativistic treatment of baryons
at low energies,  
encountered in
\cite{gasser}.
Jenkins and Manohar suggested to take extremally non-relativistic limit of 
the fully relativistic theory and expand in inverse powers of baryon mass
$M$.
    
In the heavy baryon approach one 
integrates out heavy degrees of freedom and expands resulting non-local
operators in inverse powers of large mass.  In terms of the relativistic
perturbation theory of  
the original field theoretical model (Feynman diagrams) heavy baryon
approach corresponds to the expansion  
of integrands in the loop integrals in powers of $1/M$ with subsequent term by
term integration of the resulting series \cite{ellis}. The non-commutativity of
the integration over loop momenta and the  
expansion in $1/M$ generates a problem  of matching of heavy baryon approach  to
the original relativistic theory. According Lepage's
\cite{lepage} argument from the uncertainty principle, one should be able to
compensate the difference between the 
results of ``naive'' heavy baryon and relativistic approaches by including
additional terms into the Lagrangian of the heavy baryon approach.  
While the problem of this  matching has been analysed at one loop level
\cite{ellis,bernard92,finkemeier,lutz}, to the best of our
knowledge the matching procedure for higher order loops has not been studied.

In the present paper we consider the matching problem on a two loop level on the
example of  
the forward scattering amplitude in a scalar theory. The consideration of the
non-zero  
spin and the non-zero transferred momentum makes calculations more tedious and
less 
transparent, bringing nothing new and essential in the problem considered. In
our calculations 
we use the technique of calculation of loop integrals by dimensional counting,
developed in  \cite{gjt}.

We explicitly show that heavy baryon
approach reproduces the results of the original relativistic theory at two-loop
level.   

\section{One-loop analysis}

Let us consider a field theoretical model described by the Lagrangian:

\begin{equation}
{\cal L}=-{1\over 2}\Phi^*\partial_{\mu}\partial^{\mu}\Phi 
-{1\over 2}M^2\Phi^*\Phi
-{1\over 2}\phi\partial_{\mu}\partial^{\mu}\phi - 
g\Phi^*\Phi\phi +L_1
\label{lagrangian}
\end{equation}
where $\Phi$ is a complex scalar field with mass $M$,  $\phi$ is a neutral massless
scalar
field, $g$ is a coupling constant and $L_1$ contains all counter-terms which
are necessary to remove
divergences (one can include also interactions with the derivatives and/or
a larger number of fields and corresponding counter-terms).

To avoid
complications due to the infrared singularities we work in six dimensional
space-time. We use dimensional regularization and $n$ is a dimension of
space-time.

Heavy baryon approach to the processes which involve one heavy particle uses
following expansion of the heavy scalar propagator 
($p_{\mu}=Mv_{\mu}+k_{\mu}$, $v^2=1$):
\begin{equation}
{1\over p^2-M^2}={1\over 2Mv\cdot k+k^2}={1\over 2M}{1\over v\cdot k+{k^2\over
2M}}={1\over 2M}\left( {1\over v\cdot k}-{1\over 2M}{k^2\over (v\cdot k)^2}
+\cdots\right)
\label{hbep}
\end{equation}
This expansion corresponds to the following Lagrangian:
\begin{equation}
{\cal L}_{1}=-{M}\psi^*\left( v\cdot\partial +{\partial^2\over 2M}\right)\psi
\label{hbflagrangian}
\end{equation}
where the second term of ${\cal L}_1$ is  treated perturbatively.
This Lagrangian can be obtained from the free part of ${\cal L}$ corresponding
to heavy scalar field,  
defining $\Phi=exp\{iMv\cdot x\}\psi$.
    
Let us start with the one loop self-energy correction to the scattering process
in 
the original relativistic theory, depicted
in FIG.1 a) 
(The solid line corresponds to the heavy scalar and dashed line corresponds to
massless scalar).
The expression for this  diagram is proportional to the following integral:
\be
J_{11}=\int {d^nq\over \left[ q^2+i\epsilon\right]\left[ 
(p+q)^2-M^2+i\epsilon\right]}
\label{relint1}
\ee
The straightforward integration yields 
($p'=Mv$, $p=Mv+l$, $v^2=1$, $l^2=0$, $p^2=M^2+2Mv\cdot l$):
$$
J_{11}={i{\pi }^{n\over 2}\over M}\Gamma\left( {n\over 2}-1\right)
\Gamma\left( 3-n\right)(-2v\cdot l)^{n-3}
{ }_2F_1\left( {n\over 2}-1, n-2;n-2;-{2v\cdot l\over M} \right)
$$
\begin{equation}
+i{\pi }^{n\over 2}\left( M^2\right)^{{n\over 2}-2}
{\Gamma\left( 2-
{n\over 2}\right)\Gamma\left( n-3\right)\over 
\Gamma\left( n-2\right)}{ }_2F_1\left( 1,2-n/2;4-n; -{2v\cdot l\over M}\right)
\label{relint}
\end{equation}
where ${ }_pF_q\left( a_1,...,a_p;b_1,...,b_q;z\right)$ are the
(generalised) Hypergeometric 
functions of $z$ \cite{abramowitz}. 

\begin{figure}[t]
\hspace*{1.5cm}  \epsfxsize=14cm\epsfbox{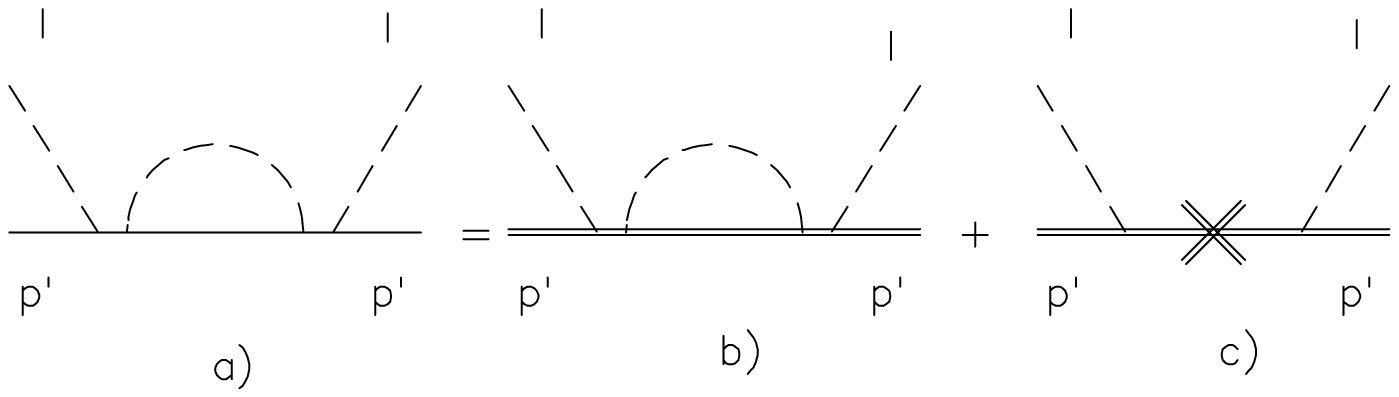}
\vspace{2mm}
\caption{\fign{se}{\it One-loop corrections to the heavy scalar propagator
in
the heavy scalar-massless scalar scattering process.}}
\end{figure}

Heavy baryon expansion for $J_{11}$ is realized by
expanding (according to Eq. (\ref{hbep})) the integrand  
in $1/M$ and integrating the resulting series term-by-term.

As it was observed in \cite{gjt},  
expanding integrand in powers of some parameter
and changing the order of integration and summation one recovers that part of
the value 
of the integral which can be expanded in powers of given parameter with non-zero
coefficients.  

From the expression (\ref{relint})  we see that the first term can be expanded
in powers of $1/M$
with non-zero coefficients, while the second one can not - it contains 
$\left( M^2\right)^{{n\over 2}-2}$.
Hence we expect that heavy baryon approach
reproduces the first term of the expression (\ref{relint}). 

Indeed:
$$
J_{11}=\int {d^nq\over \left[ q^2+i\epsilon\right]\left[ 
(p+q)^2-M^2+i\epsilon\right]}=\int {d^nq\over \left[ q^2+i\epsilon\right]\left[ 
(l+q)^2+2Mv\cdot (l+q)+i\epsilon\right]}
$$
$$
=
{1\over 2M}\int {d^nq\over \left[ q^2+i\epsilon\right]\left[ 
v\cdot (l+q)+{(l+q)^2\over 2M}+i\epsilon\right]}
$$
expanding the integrand and changing the order of integration and summation we
obtain:
$$
J_{11HB}=
{1\over 2M}\left\{ \int {d^nq\over \left[ q^2+i\epsilon\right]\left[ 
v\cdot (l+q)+i\epsilon\right]}-{1\over 2M}\int {d^nq(q+l)^2\over \left[ q^2
+i\epsilon\right]\left[ v\cdot (l+q)+i\epsilon\right]^2}\right\}+\cdots
$$ 
\begin{equation}
={i{\pi }^{n\over 2}\over M}\Gamma\left( {n\over 2}-1\right)
\Gamma\left( 3-n\right)(-2v\cdot l)^{n-3}
+{i{\pi }^{n\over 2}\over 2M^2}
\Gamma\left( {n\over 2}-1\right)\Gamma\left( 3-n\right)(n-2)(-2v\cdot l)^{n-2}+
\cdots
\label{hbint}
\end{equation}
As it was expected, Eq. (\ref{hbint}) reproduces the expansion of the first term
of the expression (\ref{relint}).

The second term of the Eq. (\ref{relint}) which can not be expanded in powers of
$1/M$ is analytic in momentum $l$ and hence can be reproduced by introducing
additional terms into the  Lagrangian of the heavy baryon approach.  

Free propagators of the heavy scalar particle appearing in the expression for
the considered diagram are apparently reproduced by heavy baryon approach.
The same is true for all diagrams and  we will not include the contributions
of the free propagators in our analysis below.

FIG.1 b) schematically represents the first term in right hand side of
the Eq. (\ref{relint}) (or Eq.(\ref{hbint})) FIG.1 c) corresponds to
the contributions of compensating terms.

\begin{figure}[t]
\hspace*{1.5cm}  \epsfxsize=14cm\epsfbox{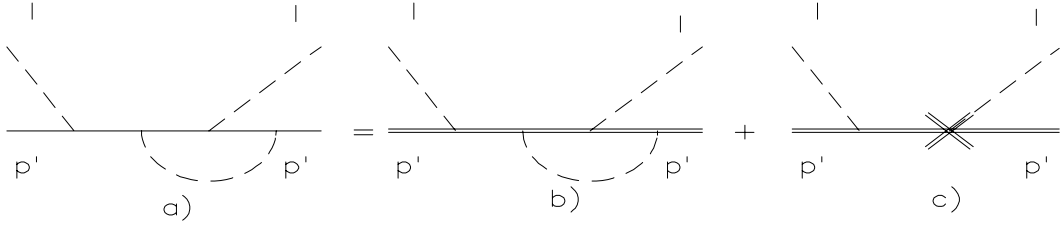}
\vspace{2mm}
\caption{\fign{sev}{\it One-loop corrections to the vertex in
the heavy scalar-massless scalar scattering process.}}
\end{figure}

Next one-loop diagram we are considering here is drawn in FIG.2 a) (one-loop
vortex correction to the light scalar-heavy scalar vertex). The result of this
diagram is proportional to the following integral:
$$
J_{12}=\int {d^nq\over \left[ q^2+i\epsilon\right]\left[ 
q^2+2p'q+i\epsilon\right]\left[ 
(p+q)^2-M^2+i\epsilon\right]}
$$
$$
=-i\left( M^2\right)^{{n\over 2}-3}{\pi }^{n\over 2}
{\Gamma\left( n-4\right)\Gamma\left( 3-n/2\right)\over 
\Gamma\left( n-3\right)}{ }_3F_2\left( 1,1,3-n/2;2,5-n;{M^2-p^2\over M^2}\right)
$$
$$
-i\left( M^2\right)^{{n\over 2}-3}{\pi }^{n\over 2}\left( 
{M^2-p^2\over M^2}\right)^{n-4}{\Gamma\left( 4-n\right)
\Gamma\left( n/2-1\right)\over n-3}
{ }_3F_2\left( n/2-1,n-3;n-2;{M^2-p^2\over M^2}\right)
$$
$$
={i{\pi }^{n\over 2}\over M^2}\Gamma\left( 3-n\right)\Gamma\left( n/2-1\right)
{(-2v\cdot l)^{n-4}}-{i{\pi }^{n\over 2}\over 2M^3}
\Gamma\left( 4-n\right)\Gamma\left( n/2-1\right){(-2v\cdot l)^{n-3}}+
\cdots 
$$
\be
-i\left( M^2\right)^{{n\over 2}-3}{\pi }^{n\over 2}
{\Gamma\left( n-4\right)\Gamma\left( 3-n/2\right)\over 
\Gamma\left( n-3\right)}
{ }_3F_2\left( 1,1,3-n/2;2,5-n;{-2v\cdot l\over M}\right)
\label{3}
\end{equation}
On the other hand heavy baryon approach leads to:
$$
J_{12HB}=
{1\over 4M^2}\int {d^nq\over \left[ q^2+i\epsilon\right]\left[ 
v\cdot (l+q)+i\epsilon\right]\left[ 
v\cdot q+i\epsilon\right]}
$$
$$
-{1\over 8M^3}\int {d^nq(q+l)^2\over \left[ q^2
+i\epsilon\right]\left[ v\cdot (l+q)+i\epsilon\right]^2\left[ 
v\cdot q+i\epsilon\right]}+\cdots
$$ 
\begin{equation}
={i{\pi }^{n\over 2}\over M^2}\Gamma\left( {n\over 2}-1\right)
\Gamma\left( 3-n\right)(-2v\cdot l)^{n-4}
-{i{\pi }^{n\over 2}\over 2M^3}
\Gamma\left( {n\over 2}-1\right)\Gamma\left( 4-n\right)(-2v\cdot l)^{n-3}+
\cdots
\label{hbint1}
\end{equation}

The comparison of Eq. (\ref{3}) and Eq. (\ref{hbint1}) shows that heavy baryon
approach reproduces that part of relativistic answer 
which can be expanded in inverse powers of $M$.
The second part is analytic in $l$ and can be reproduced by adding appropriate
terms into the Lagrangian of the heavy baryon approach. FIG.2 b) and FIG.2 c)
correspond to Eq. (\ref{hbint1}) and the contributions of compensating terms
respectively. 

The analysis of the rest of one-loop diagrams lead to the same result:
heavy baryon
approach reproduces those parts of diagrams which 
are non-analytic 
in the momenta and the remaining  parts, analytic in momenta can be reproduced
by 
adding terms into the effective Lagrangian of the heavy baryon approach.

\section{Two-loop analysis}

Two-loop diagrams have more complicated structure. Let us consider two-loop
correction to the propagator of the heavy scalar in original relativistic theory
depicted in FIG.3 a). The result of
this diagram is proportional to the following integral:

\begin{equation}
J_{21}=\int {d^nq_1d^nq_2\over \left[ q_1^2+i\epsilon\right]
\left[ q_2^2+i\epsilon\right]
\left[ \left( p+q_1\right)^2-M^2+i\epsilon \right]^2
\left[ \left( p+q_1+q_2\right)^2-M^2+i\epsilon \right]}
\label{rainbow}
\end{equation}
From the method of dimensional counting \cite{gjt} it follows that
\be
J_{21}=\delta^{2n-7}\left( p^2\right)^{n-5}\sum_{k=0}^{\infty}f_{1k}\delta^k
+\delta^{n-4}\left( p^2\right)^{n-5}\sum_{k=0}^{\infty}f_{2k}\delta^k+
\left( p^2\right)^{n-5}\sum_{k=0}^{\infty}f_{3k}\delta^k
\label{expansion}
\ee
where
\be
\delta ={M^2-p^2\over p^2}
={-2v\cdot l\over M}{1\over 1+{2v\cdot l\over M}}
\label{delta}
\ee
and the coefficients $f_{ik}$ are determined by the original integral
(\ref{rainbow}) \cite{gjt}.
Substituting Eq. (\ref{delta}) into Eq. (\ref{expansion}) we obtain:
$$
J_{21}=\left( -2v\cdot l\right)^{2n-7}M^{-3}\sum_{k=0}^{\infty}D_{1k}
\left( {2v\cdot l\over M}\right)^k
+\left( -2v\cdot l\right)^{n-4}M^{n-6}\sum_{k=0}^{\infty}D_{2k}
\left( {2v\cdot l\over M}\right)^k
$$
\be
+M^{2n-10}\sum_{k=0}^{\infty}D_{3k}\left( {2v\cdot l\over M}\right)^k
\label{j2exp}
\ee
where $D_{ik}$ do not depend on $M$, $l$ or $v$.

\begin{figure}[t]
\hspace*{3cm}  \epsfxsize=12cm\epsfbox{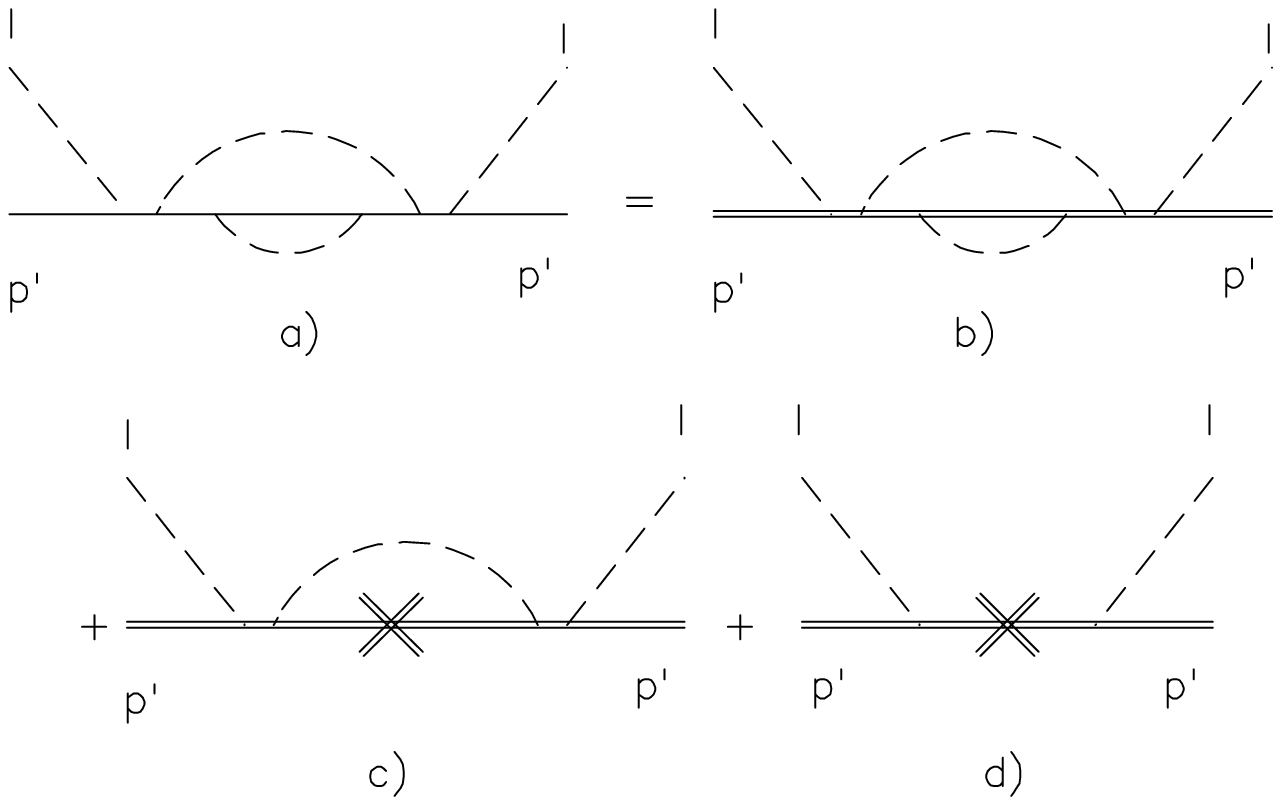}
\vspace{2mm}
\caption{\fign{ser}{\it Two-loop correction to the heavy scalar propagator in
the heavy scalar-massless scalar scattering process.}}
\end{figure}

Actual calculations of $J_{21}$ can be performed using the methods of \cite{gjt}
as follows. 
Let us rewrite:

\begin{equation}
J_{21}=\int {d^nq_1d^nq_2\over \left[ q_1^2+i\epsilon\right]
\left[ q_2^2+i\epsilon\right]
\left[ 2Mv\cdot \left( l+q_1\right)+\left( l+q_1\right)^2+i\epsilon \right]^2
\left[ 2Mv\cdot \left( l+q_1+q_2\right)+\left( l+q_1+q_2\right)^2+i\epsilon
\right]}
\label{rainbowvl}
\end{equation}
First we expand the integrand in inverse powers of $M$,  change
the order of integration and summation and obtain:
$$
J_{21}^{(1)}={1\over 4M^3}\int {d^nq_1d^nq_2\over \left[ q_1^2+i\epsilon\right]
\left[ q_2^2+i\epsilon\right]
\left[ v\cdot \left( l+q_1\right)+i\epsilon \right]^2
\left[ v\cdot \left( l+q_1+q_2\right)+i\epsilon \right]}
$$
$$
-{1\over 4M^4}\Biggl (\int {d^nq_1d^nq_2 \ \left( q_1+l\right)^2
\over \left[ q_1^2+i\epsilon\right]
\left[ q_2^2+i\epsilon\right]
\left[ v\cdot \left( l+q_1\right)+i\epsilon \right]^3
\left[ v\cdot \left( l+q_1+q_2\right)+i\epsilon \right]}
$$
\begin{equation}
+{1\over 2}\int {d^nq_1d^nq_2 \ \left( q_1+q_2+l\right)^2
\over \left[ q_1^2+i\epsilon\right]
\left[ q_2^2+i\epsilon\right]
\left[ v\cdot \left( l+q_1\right)+i\epsilon \right]^2
\left[ v\cdot \left( l+q_1+q_2\right)+i\epsilon \right]^2}\Biggr) +\cdots
\label{rainbowvl1}
\end{equation}

Second we re-scale $q_2\rightarrow q_2M$, extract a non-integer power of mass,
expand the 
integrand in inverse powers and change the order of integration and summation.
The result is:
$$
J_{21}^{(2)}={M^{n-6}\over 4}\int {d^nq_1d^nq_2\over \left[ q_1^2
+i\epsilon\right]
\left[ q_2^2+i\epsilon\right]
\left[ v\cdot \left( l+q_1\right)+i\epsilon \right]^2
\left[ q_2^2+2v\cdot q_2+i\epsilon \right]}
$$
$$
-{M^{n-7}\over 4}\Biggl (\int {d^nq_1d^nq_2 \ \left( q_1+l\right)^2
\over \left[ q_1^2+i\epsilon\right]
\left[ q_2^2+i\epsilon\right]
\left[ v\cdot \left( l+q_1\right)+i\epsilon \right]^3
\left[ q_2^2+2v\cdot q_2+i\epsilon \right]}
$$
\begin{equation}
+\int {d^nq_1d^nq_2 \ \left\{ v\cdot\left( q_1+l\right)-
2q_2\cdot\left( q_1+l\right)\right\}
\over \left[ q_1^2+i\epsilon\right]
\left[ q_2^2+i\epsilon\right]
\left[ v\cdot \left( l+q_1\right)+i\epsilon \right]^2
\left[ q_2^2+2v\cdot q_2+i\epsilon \right]^2}\Biggr)+\cdots
\label{rainbowvl2}
\end{equation} 
and third we re-scale $q_1\rightarrow Mq_1$, $q_2\rightarrow Mq_2$, extract
non-integer power of the 
mass, change the order of integration and summation and obtain:
$$
J_{21}^{(3)}=M^{2n-10}\int {d^nq_1d^nq_2\over \left[ q_1^2+i\epsilon\right]
\left[ q_2^2+i\epsilon\right]
\left[ q_1^2+2v\cdot q_1+i\epsilon \right]^2
\left[ \left( q_1+q_2\right)^2+2v\cdot \left( q_1+q_2\right)+i\epsilon \right]}
$$
$$
-M^{2n-11}\Biggl (\int {d^nq_1d^nq_2 \ \left( 4l\cdot q_1+4v\cdot l\right)
\over \left[ q_1^2+i\epsilon\right]
\left[ q_2^2+i\epsilon\right]
\left[ q_1^2+2v\cdot q_1+i\epsilon \right]^3
\left[ \left( q_1+q_2\right)^2+2v\cdot \left( q_1+q_2\right)+i\epsilon \right]}
$$
\begin{equation}
+\int {d^nq_1d^nq_2 \ \left\{ 2l\cdot\left( q_1+q_2\right)+
2v\cdot l\right\}
\over \left[ q_1^2+i\epsilon\right]
\left[ q_2^2+i\epsilon\right]
\left[ q_1^2+2v\cdot q_1+i\epsilon \right]^2
\left[ \left( q_1+q_2\right)^2+2v\cdot \left( q_1+q_2\right)
+i\epsilon \right]^2}\Biggr)+\cdots
\label{rainbowvl3}
\end{equation} 
Integral $J_{21}$ is nothing else than a sum of $J_{21}^{(1)}$, $J_{21}^{(2)}$
and
$J_{21}^{(3)}$ \cite{gjt}.

Evidently, $J_{21}^{(1)}$ is expandable
in inverse powers of $M$ and hence heavy baryon
approach  reproduces
this part of $J_{21}$.  $J_{21}^{(3)}$
can not be reproduced (because it contains $M^{2n-10}$) but it is analytic in
$l$ and hence it can be taken into account
by adding compensating terms into the Lagrangian of the heavy baryon
approach. It is $J_{21}^{(2)}$, corresponding to the second term in
Eq. (\ref{j2exp}) which 
is non-trivial and might cause problems: there can
appear terms which are not expandable in powers of 
$M$ and have non-analytic dependence on $l$. 

This feature does not appear on a one loop level.  
Let us consider this problem in
details. 


The above given representation for $J_{21}$ can be obtained also as follows:
One loop sub-integral over $q_2$ can be represented as a sum of two parts: the
first part is a result of expanding integrand of this sub-integral in inverse
powers of $M$ and changing the order of integration and summation. The second
part is obtained by rescaling $q_2\rightarrow q_2M$, extracting non-integer
factor of $M$, 
expanding the integrand in powers of $1/M$ and changing the order of
integration and summation:
\be
J_{21}=\int {d^nq_1\over \left[ q_1^2+i\epsilon\right]
\left[ 2Mv\cdot \left( l+q_1\right)+\left( l+q_1\right)^2
+i\epsilon \right]^2}\left\{ 
F_1\left( M,l+q_1\right)+M^{n-4}F_2\left( M,l+q_1\right)\right\}
\label{subexp}
\ee
where $F_1$ and $F_2$ represent series in $1/M$. As we concluded
from one-loop analysis heavy baryon approach reproduces $F_1$ at one loop
level and $M^{n-4}F_2$ is reproduced by adding compensating terms into the 
Lagrangian of the heavy baryon approach. 

Now, expanding the denominator appearing in Eq. (17)
$$
{1\over \left[ q_1^2+i\epsilon\right]
\left[ 2Mv\cdot \left( l+q_1\right)+\left( l+q_1\right)^2
+i\epsilon \right]^2}
$$
in inverse powers of $M$ and changing the order of integration and summation in
Eq. (\ref{subexp}) we get the result which is equal to
$J_{21}^{(1)}+J_{21}^{(2)}$. 
This makes clear that heavy baryon approach reproduces $J_{21}^{(2)}$ which was
addressed 
as a possible source of the trouble. As for $J_{21}^{(3)}$ it is analytic in $l$
and can be reproduced by compensating terms. Note that $J_{21}^{(3)}$ is
obtained from Eq. (\ref{subexp}) by
rescaling $q_1\rightarrow q_1M$: one extracts non-integer power of $M$, expands
the 
integrand in powers of $1/M$ and changes the order of integration and
summation. Doing so one gets $M^{n-6}f_1+M^{2n-10}f_2$, where $M^{2n-10}f_2$ is
equal to $J_{21}^{(3)}$ and $f_1$ turns out to be a sum of trivial terms
(zeros). 

FIG.3 b), FIG.3 c) and FIG.3 d) correspond to $J_{21}^{(1)}$, $J_{21}^{(2)}$ and
$J_{21}^{(3)}$ respectively.

\begin{figure}[t]
\hspace*{1cm}  \epsfxsize=14cm\epsfbox{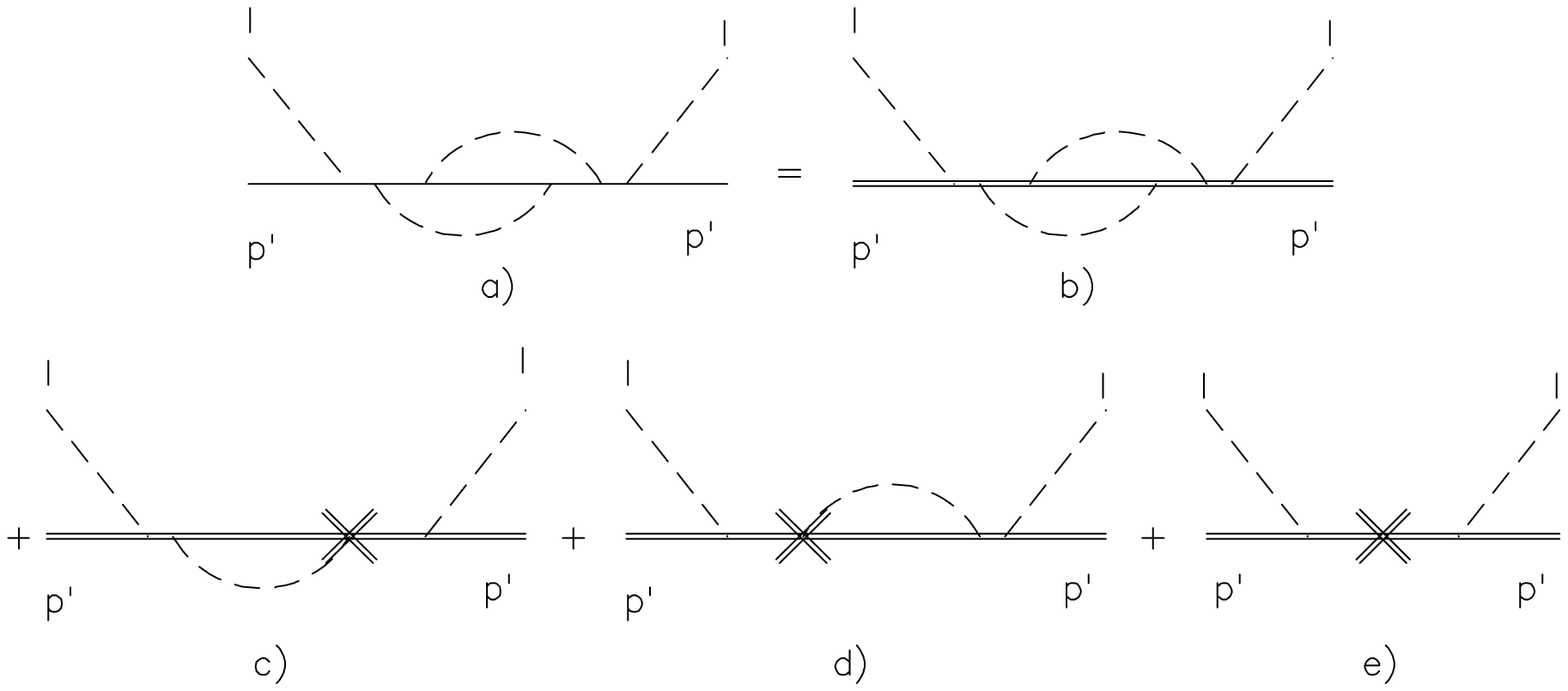}
\vspace{2mm}
\caption{\fign{sep}{\it Two-loop correction to the heavy scalar propagator in
the heavy scalar-massless scalar scattering process.}}
\end{figure}

Next we consider two-loop correction to the heavy scalar propagator in original
relativistic theory depicted in
FIG.4. The result of this diagram is proportional to the following integral:

\begin{equation}
J_{22}=\int {d^nq_1d^nq_2\over \left[ q_1^2+i\epsilon\right]
\left[ q_2^2+i\epsilon\right]
\left[ \left( p+q_1\right)^2-M^2+i\epsilon \right]
\left[ \left( p+q_1+q_2\right)^2-M^2+i\epsilon \right]
\left[ \left( p+q_2\right)^2-M^2+i\epsilon \right]}
\label{perekriostka}
\end{equation}

For later use before analysing Eq. (\ref{perekriostka}) let us consider $J_v$ -
an off-mass 
shell integral of the one loop correction to the vertex: 
\begin{equation}
J_v=\int {d^nq\over \left[ q^2+i\epsilon\right]\left[ 
(p+q)^2-M^2+i\epsilon \right]\left[ 
(k'+q)^2-M^2+i\epsilon\right]}
\label{vertexint0}
\end{equation}
where $p=Mv+l$ and $k'=Mv+l'$.
\begin{equation}
J_v=\int {d^nq\over \left[ 2Mv\cdot (l+q)+(l+q)^2+i\epsilon \right] 
\left[ 2Mv\cdot (l'+q)+(l'+q)^2+i\epsilon \right]}=J_v^{1}(l,l')+J_v^{2}(l,l') 
\label{vertexint1}
\end{equation}
where $J_v^{1}$ is obtained by expanding the integrand of
Eq. (\ref{vertexint1}) in
inverse powers of $M$ and changing the order of integration and summation:
$$
J_v^{1}(l,l')={1\over 4M^2}\int {d^nq\over \left[ q^2+i\epsilon \right]
\left[ v\cdot l+v\cdot q+i\epsilon \right] 
\left[ v\cdot l'+v\cdot q+i\epsilon \right]}  
$$
$$
-{1\over 8M^3}\Biggl (\int {d^nq\over \left[ q^2+i\epsilon \right]
\left[ v\cdot l+v\cdot q+i\epsilon \right]^2 
\left[ v\cdot l'+v\cdot q+i\epsilon \right]}  
$$
\be
+\int {d^nq\over \left[ q^2+i\epsilon \right]
\left[ v\cdot l+v\cdot q+i\epsilon \right] 
\left[ v\cdot l'+v\cdot q+i\epsilon \right]^2}\Biggr)+\cdots  
\label{vertexint2}
\ee

$J_v^2$ is obtained by rescaling $q\rightarrow qM$, extracting
non-integer power of $M$, expanding the
integrand in negative powers of the mass and changing the order of integration
and summation:
$$
J_v^2(l,l')=M^{n-6}\int {d^nq\over \left[ q^2+i\epsilon \right]
\left[ 2v\cdot q+q^2+i\epsilon \right]^2}  
-M^{n-7}\Biggl (\int {d^nq \ 2(v\cdot l+l\cdot q)\over \left[ q^2+i\epsilon
\right] 
\left[ 2v\cdot q+q^2+i\epsilon \right]^3}  
$$
\be
+\int {d^nq 2(v\cdot l'+l'\cdot q)\over \left[ q^2+i\epsilon \right]
\left[ 2v\cdot q+q^2+i\epsilon \right]^3}\Biggr)+\cdots  
\label{vertexint3}
\ee
Heavy baryon approach reproduces $J_v^1$. $J_v^2$ is analytic in $l$ and $l'$
and can be reproduced by adding compensating terms into the Lagrangian of the
heavy baryon approach. 

\medskip

Applying the method of dimensional counting \cite {gjt} to $J_{22}$ we
obtain the following expression:
$$
J_{22}=\left( -2v\cdot l\right)^{2n-7}M^{-3}\sum_{k=0}^{\infty}A_{1k}
\left( {2v\cdot l\over M}\right)^k
+\left( -2v\cdot l\right)^{n-3}M^{n-7}\sum_{k=0}^{\infty}A_{2k}
\left( {2v\cdot l\over M}\right)^k
$$
\be
+\left( -2v\cdot l\right)^{n-3}M^{n-7}\sum_{k=0}^{\infty}A_{3k}
\left( {2v\cdot l\over M}\right)^k
+M^{2n-10}\sum_{k=0}^{\infty}A_{4k}\left( {2v\cdot l\over M}\right)^k
=J_{22}^{(1)}+J_{22}^{(2)}+J_{22}^{(3)}+J_{22}^{(4)}
\label{dimcj3}
\ee
where $A_{ik}$ do not depend on $M$, $l$ or $v$. In Eq. (\ref{dimcj3})
$J_{22}^{(1)}$ 
is the result of expanding integrand in inverse powers of $M$ and integrating
the series, 
$J_{22}^{(2)}$ and $J_{22}^{(3)}$ are obtained by rescaling $q_1\rightarrow
q_1M$ and 
$q_2\rightarrow q_2M$ 
correspondingly with subsequent expansion of the integrand and change of the
order 
of integration and summation and $J_{22}^{(4)}$ is the result of the 
simultaneous 
rescaling $q_1\rightarrow q_1M$ and $q_2\rightarrow q_2M$ and integration of the
resulting expansion. 

Heavy baryon
approach reproduces straightforwardly  $J_{22}^{(1)}$;  
$J_{22}^{(4)}$ is analytic in momenta and hence can be reproduced by
compensating terms in the Lagrangian of the heavy baryon approach.  The terms
$J_{22}^{(2)}$ 
and 
$J_{22}^{(3)}$ are not expandable in $1/M$ and they are
not analytic in momenta. In a full analogy with the previous analysis for
$J_{21}^{2}$  
these terms are reproduced 
by taking into account the contributions of the compensating
terms which have to
be introduced into the Lagrangian of the heavy baryon approach in order to
reproduce the 
expression for the one loop sub-diagrams of this two-loop diagram.

To see that this is actually the case let us represent $J_{22}$ in the following
way: 
$$
J_{22}=\int {d^nq_1\over \left[ q_1^2+i\epsilon\right]
\left[ \left( p+q_1\right)^2-M^2+i\epsilon \right]}
\int {d^nq_2\over 
\left[ q_2^2+i\epsilon\right]
\left[ \left( p+q_1+q_2\right)^2-M^2+i\epsilon \right]
\left[ \left( p+q_2\right)^2-M^2+i\epsilon \right]}
$$
\be
\int {d^nq_1\over \left[ q_1^2+i\epsilon\right]
\left[ 2Mv\cdot\left( l+q_1\right)+\left( l+q_1\right)^2+i\epsilon \right]}
\left\{ J_v^{1}(l+q_1,l)+J_v^{2}(l+q_1,l)\right\} 
\label{perekriostka1}
\end{equation}
Note that $J_v^{2}(l+q_1,l)$ corresponds to 
compensating terms included into the Lagrangian of the heavy baryon approach.
 
Expanding denominator 
\be
{1\over \left[ q_1^2+i\epsilon\right]
\left[ 2Mv\cdot\left( l+q_1\right)+\left( l+q_1\right)^2+i\epsilon \right]}
\label{integrand2}
\ee
in $1/M$ and changing the order of integration and summation we reproduce 
$J_{22}^{(1)}+J_{22}^{(2)}$. So, heavy baryon approach reproduces these two
terms (here $J_{22}^{(2)}$ occurred because we included contributions of 
compensating terms corresponding to one-loop sub-diagrams).  

$J_{22}^{(3)}$ and $J_{22}^{(4)}$ are reproduced by rescaling $q_1\rightarrow
q_1M$, extracting 
non-integer factors of $M$, expanding integrand in $1/M$
and changing the order of integration and summation.

$J_{22}^{(4)}$ is analytic in momenta and hence can be reproduced by
compensating 
terms included into the Lagrangian of the heavy baryon approach. As for
$J_{22}^{(3)}$ it is equal to  $J_{22}^{(2)}$ and comes from one-loop
compensating 
terms as well. This fact should be clear from FIG.4 where FIG.4 b), FIG.4 c),
FIG.4 d) and FIG.4 e)
correspond to $J_{22}^{(1)}$, $J_{22}^{(2)}$, $J_{22}^{(3)}$ and
$J_{22}^{(4)}$ 
respectively. 

Analogous results are obtained for all the remaining two-loop diagrams.

From the above analysis it follows that heavy
baryon approach
reproduces the results of the original relativistic theory at a two-loop
order.

\section{Conclusions}
In this work we have addressed the problem of matching of heavy baryon approach
to 
the original relativistic theory. The heavy Baryon approach corresponds to
the expansion of the integrand in inverse powers of the large mass with
subsequent change of the order of integration and summation. As this two
procedures are not commutative, the difference has to be
compensated by adding terms into the Lagrangian of the heavy baryon
approach. As the addressed problem does not actually depend on the details of
the given model we considered a simple example of heavy and massless interacting
scalar fields. Using the method of calculation of loop integrals by
dimensional counting outlined in \cite{gjt} we analysed one and two loop
diagrams and demonstrated how the difference between relativistic and heavy
baryon calculations is compensated by adding terms to the Lagrangian of
the heavy baryon approach. At two-loop level the
difference can be compensated only after one includes the contributions of
compensating terms for one loop sub-diagrams. While we included only selected
diagrams in this paper, the very same conclusions
are valid for one and two loop diagrams which were not included in here.

We believe that the iterative procedure of considering contributions of
compensating terms for one-loop diagrams   
in two-loop calculations which is crucial to resolve the matching 
problem leads to analogous results for higher loops.

While we considered a simple model of scalar fields the problems of
interchange of integration and expansion in inverse powers
of heavy particle mass are the same for more realistic models with included
fermionic 
and vector fields. 
In heavy baryon chiral perturbation theory the compensating terms with similar
structure are actually summed up and included as redefinitions of already
existing coupling constants. This redefinition is crucial, it actually leads to
the consistent power counting of the heavy baryon chiral perturbation theory. In
heavy baryon 
approach the coupling constants which correspond to re-defined relativistic
coupling constants are introduced as initially free parameters which are to be
fixed from experimental data. Working up to some given order in heavy baryon
approach 
one actually re-sums low order contributions of an infinite number of
relativistic high order loop diagrams.    

It was shown in Ref. \cite{bernard} that, within HBCHPT, an infinite number of
internal line insertions must be summed to describe the scalar form-factor of
the nucleon near threshold. As we demonstrated above the heavy baryon expansion
reproduces the expansion of the relativistic result. This conclusion is formally
still correct for the scalar form-factor of the nucleon, but the problem is that
the expansion of the relativistic result 
is not convergent near threshold. This problem has been successfully resolved
recently by Becher and Leutwyler using ``infrared regularization''
\cite{becher}.  

As was demonstrated above relativistic diagrams contain parts which can not be
altered by adding local terms into the Lagrangian. These parts are directly
reproduced by heavy baryon approach and they respect
power counting.  Other parts which are responsible for violation
of the power 
counting in relativistic theory can be changed by adding counter-terms.
Hence it should be more or less clear that the
problems of the relativistic approach, in particular that multi-loop diagrams
contribute into low order calculations encountered in \cite{gasser}, can be
solved within relativistic approach using appropriately chosen normalisation
condition. Hence one could from the very beginning work within original
relativistic approach and never encounter the problems of near threshold
behaviour of the scalar form-factor of the nucleon. These problems will be
addressed in next paper.

\medskip
\medskip

{\bf ACKNOWLEDGEMENTS} 

This work was carried out whilst one of the authors (J.G.) was a recipient of 
an Overseas Postgraduate
Research Scholarship and a Flinders University Research Scholarship
holder at Flinders University of South Australia.

This work was supported in part by National Science Foundation under 
Grant No. HRD9450386, Air Force Office of Scientific Research under 
Grant No. F4962-96-1-0211 and Army Research Office under Grant No. 
DAAH04-95-1-0651.

\end{document}